\documentclass[12pt]{article}

\usepackage{graphics}

\def\eq#1{Eq.~(\ref{#1})}
\newcommand{\secn}[1]{Section~\ref{#1}}
\newcommand{\bra}[1]{\langle{#1}|}
\newcommand{\ket}[1]{|{#1}\rangle}

\newcommand{\nl}{\nonumber \\}

\def\beq{\begin{equation}}
\def\eeq{\end{equation}}
\def\beqa{\begin{eqnarray}}
\def\eeqa{\end{eqnarray}}

\newcommand{\sect}[1]{\setcounter{equation}{0}\section{#1}}

\newcommand{\as}{\alpha_s}
\renewcommand{\b}{\beta}

\newcommand{\e}{\epsilon}

\renewcommand{\thefootnote}{\fnsymbol{footnote}}

\def\bet{\begin{tabular}}
\def\eet{\end{tabular}}

\begin{document}
\begin{titlepage}
\rightline{DFTT 22/00}
\rightline{\hfill June 2000}

\vskip 3cm

\centerline{\Large \bf Analytic resummation}
\centerline{\Large \bf for the quark form factor in QCD} 
 
\vskip 2cm

\centerline{\bf Lorenzo Magnea\footnote{e-mail: magnea@to.infn.it}}
\centerline{\sl Dipartimento di Fisica Teorica, Universit\`a di Torino}
\centerline{\sl and I.N.F.N., Sezione di Torino}
\centerline{\sl Via P.Giuria 1, I--10125 Torino, Italy}

 \vskip 2cm
 
\begin{abstract}

The quark form factor is known to exponentiate within the framework of
dimensionally regularized perturbative QCD. The logarithm of the form
factor is expressed in terms of integrals over the scale of the
running coupling.  I show that these integrals can be evaluated
explicitly and expressed in terms of renormalization group invariant
analytic functions of the coupling and of the space--time dimension,
to any order in renormalized perturbation theory. Explicit expressions
are given up two loops. To this order, all the infrared and collinear
singularities in the logarithm of the form factor resum to a single
pole in $\e$, whose residue is determined at one loop, plus powers of
logarithms of $\e$. This behavior is conjectured to extend to all
loops.

\end{abstract}

\end{titlepage}

\newpage
\renewcommand{\thefootnote}{\arabic{footnote}}
\setcounter{footnote}{0}
\setcounter{page}{1}

\sect{Introduction}
\label{intro}

The asymptotic behavior of the electromagnetic form factor of charged
particles at high energy has been the object of theoretical studies
for almost half a century. It is the simplest gauge theory amplitude
to exhibit a doubly logarithmic behavior, as a consequence of its
infrared and collinear singularities. Thus the understanding of its
asymptotic energy dependence requires a resummation of perturbation
theory, going beyond renormalization group logarithms. This
resummation was performed for the first time, at the leading logarithm
level, by Sudakov \cite{suda}, who condidered the off--shell form
factor for an abelian gauge theory. He found that the leading (double)
logarithms exponentiate, which results in the strong suppression of
the elastic scattering of charged particles at high energy which bears
his name. A similar exponentiation for the on--shell form factor was
obtained in \cite{jack}, and in later years the result was extended to
nonabelian gauge groups \cite{DDT}.

Further extending the exponentiation to all subleading logarithms is a
nontrivial task, which requires the complex machinery of perturbative
factorization theorems. This goal was achieved by Mueller \cite{muel}
and Collins \cite{coll} for abelian theories, and shortly later by Sen
\cite{sen} for QCD. In essence, the complete exponentiation is
achieved by showing that the energy dependence of the form factor must
follow an evolution equation, which embodies the constraints of
renormalization group and gauge invariance. This evolution equation
was typically solved in terms of an undetermined initial condition,
containing low--energy information and depending on some chosen
infrared regulator, say a gluon mass. It was later realized \cite{us}
that dimensional regularization can be used directly at the level of
the evolution equation. One can then give a simple and explicit
solution for the form factor, using as initial condition the electric
charge at vanishing momentum transfer. When the solution is written in
this form, all infrared and collinear poles are explicitly
exponentiated, and the results can be directly compared with
diagrammatic calculations.

The quark form factor, although divergent, is a quantity of
considerable interest for phenomenology: in fact, it appears in the
computation of several cross sections, and techniques similar to the
ones used for its exponentiation have been applied to several
processes of interest in high energy physics, in particular to
processes in which the perturbative series needs to be resummed
because of logarithmic enhancements in special regions of phase
space. Outstanding examples are the inclusive Drell--Yan cross section
\cite{dysum} and the transverse momentum distribution of vector bosons
produced in hadronic collisions \cite{ptsum}. More recently, similar
techniques have been generalized to hadronic cross section with
colored particles in the final state \cite{ttbar}. Strikingly, the
quark form factor enters directly in the hard partonic cross section
for the Drell--Yan process in the DIS scheme, which is proportional to
the modulus squared of the ratio of the timelike to the spacelike form
factor \cite{dysum,pari}. In \cite{us} it was shown that this ratio is
given by an infinite phase (a generalization of the Coulomb phase,
expressed by a series of pure counterterms in the $\overline{MS}$
scheme), times a manifestly finite exponential factor.

In the present paper, I will build upon the results of Ref.~\cite{us},
and show that the resummed expression for the form factor can be
explicitly evaluated to the desired accuracy, thus expressing the
logarithm of the form factor as an analytic function of
$\alpha_s(Q^2)$ and $\e$, manifestly renormalization group
invariant. The key observation is the following: typically, in
resummed expressions, integrals over the renormalization scale cannot
be explicitly evaluated, because the one--loop running coupling has a
Landau pole on the integration contour, which extends from the hard
scale, $\mu^2 = Q^2$, all the way down to $\mu^2 = 0$, in the region
dominated by nonperturbative effects. The sensitivity to the Landau
pole, which can be estimated using various methods and models
\cite{bene}, is then taken as a measure of the influence of
nonperturbative effects on the given observable.  In dimensional
regularization, however, the running coupling depends not only on the
renormalization scale, but also on the dimensional regularization
parameter $\e$. At one loop, the coupling still exhibits a Landau
pole, but for general (complex) $\e$ the pole is located at complex
values of the renormalization scale $\mu^2$, away from the integration
contour. As a consequence, all integrals appearing in the logarithm of
the form factor are well--defined, and can be evaluated.
Alternatively, one can express the one--loop running coupling as a
power series in terms of $\alpha_s(Q^2)$, and integrate the series
term by term: one finds that there is no factorial growth in the
coefficients of the series obtained after integration; in fact, the
series can be summed to reconstruct the same analytic function of
$\alpha_s$ and $\e$ that was found by direct integration. It is easy
to see how these results generalize when the running coupling is
expressed including higher orders in the QCD $\beta$ function. The
complexity of the answer grows, but the most relevant features remain
unchanged; in particular, the answer is manifestly invariant under
changes in the renormalization scale, to the given accuracy in the
$\beta$ function; also, infrared and collinear poles resum, so that
the logarithm of the form factor has only a single pole in $\e$, as
well as cut on the negative real axis in the $\e$ plane; conventional
perturbative results can be recovered by reexpanding in powers of
$\alpha_s$ for finite $\e$.

The outline of the paper is the following: in \secn{expon}, I briefly
review the known results on the exponentiation of the quark form
factor. In \secn{count}, I illustrate the power of dimensional
regularization by solving explicitly to all orders the recursion
relation arising from the renormalization group equation for the
generalized Coulomb phase: all the towers of poles (leading,
next--to--leading, etc.) arising in the series of counterterms can be
expressed in terms of the perturbative coefficients of the relevant
anomalous dimension, and of the QCD $\beta$ function; all the
resulting series can be summed, and they give functions of $\e$ with
at most a logarithmic singularity; the leading behavior as $\e \to 0$
can be explicitly computed to all orders. In \secn{onelo}, I turn to
the main issue, and compute the logarithm of the form factor using
one--loop results only, as well as the one--loop running coupling; the
result is remarkably simple, a single dilogarithm and a single
logarithm of functions of $\as$ and $\e$, explicitly independent of
the choice of renormalization scale. In \secn{twolo}, I generalize the
calculation to two loops, obtaining again a relatively simple answer,
containing only logarithms and dilogarithms. It is fairly clear that
the calculation could be formally pushed to even higher orders, adding
to the complexity of the answer, but without changing its basic
features; furthermore, the perturbative coefficients of the various
functions appearing in the logarithm of the form factor are known only
to two loops. Section~\ref{sumpe} summarizes the results and briefly
discusses possible applications and future developments.

\sect{Exponentiation of the form factor}
\label{expon}

The techniques needed to compute the quark form factor in perturbative
QCD to all orders in the logarithms of the energy are described in
detail in Ref.~\cite{collrev}. Here I will just give the relevant
definitions and sketch the main results, in the spirit of
Ref.~\cite{us}.

The (timelike) quark form factor in dimensionally regulated massless
QCD is defined by 
\beqa \Gamma_\mu (p_1, p_2; \mu^2, \e) & = & \bra{0}
J_\mu (0) \ket{{p_1,p_2}} \label{def} \\
& = & - {\rm i} e e_q ~\overline{v}(p_2) \gamma_\mu
u(p_1) ~\Gamma \left( \frac{Q^2}{\mu^2}, \as(\mu^2), \e \right)~, \nonumber
\eeqa
where $J_\mu (x)$ is the electromagnetic quark current, while
$\ket{{p_1,p_2}}$ is a state containing a quark and an antiquark with
momenta $p_1$ and $p_2$, and total energy $Q^2 = (p_1 + p_2)^2$, taken
to be much larger than the confinement scale. In \eq{def} I made use
of the fact that in the massless limit the perturbative form factor
has only one spin structure, and I defined the scalar form factor
$\Gamma$ to be normalized to $1$ in the absence of strong
interactions. Eq.~(\ref{def}) is written in the renormalized theory,
so that all ultraviolet divergences arising in perturbation theory
have been dealt with, and one can take $\e = 2 - d/2 < 0$ to regulate
infrared and collinear poles. Notice that, as a consequence of
electromagnetic current conservation, the anomalous dimension of the
form factor vanishes.

The conventional analysis of the form factor~\cite{collrev} proceeds
through the steps that are characteristic of the derivation of
perturbative factorization theorems: one starts by identifying the
integration regions in momentum space that contribute to the form
factor at leading power\footnote{Technically, this is done using a
massive regulator; in dimensional regularization {\it all} the energy
dependence is logarithmic, leading regions are responsible for
infrared and collinear poles, and one is in fact working with the {\it
whole} form factor.} of $Q^2$; next, one constructs a factorized
expression for the form factor, in terms of (nonlocal) operators,
typically defined in terms of eikonal lines, whose matrix elements
reproduce the leading behavior in the various regions. Each operator
factor will typically depend both on the gauge choice and on the
renormalization scale; imposing gauge and renormalization group ($RG$)
invariance leads to an evolution equation describing the energy
dependence of the whole form factor. In dimensional regularization,
this equation takes the form
\beq
Q^2 \frac{\partial}{\partial Q^2} \log \left[\Gamma \left( \frac{Q^2}{\mu^2}, 
\as(\mu^2), \e \right) \right] = \frac{1}{2} \left[ K \left(\e, \as(\mu^2) 
\right) + G \left(\frac{Q^2}{\mu^2}, \as(\mu^2), \e \right) \right]~.
\label{eveq}
\eeq
The functions $K$ and $G$ are characterized as follows: $K$ is a pure 
counterterm, thus a series of poles in the $\overline{MS}$ scheme, which I
will use throughout; $G$ contains all the energy dependence, and is finite
as $\e \to 0$; the sum $K + G$ must be renormalization group invariant, 
because the form factor is; thus $K$ and $G$ renormalize additively, 
according to
\beqa
\left( \mu \frac{\partial}{\partial \mu} + \beta(\e, \as) 
\frac{\partial}{\partial \as} \right) G \left(\frac{Q^2}{\mu^2}, \as, 
\e \right) & = & \gamma_K (\as) \label{gamk} \\ 
& = & - \left( \mu \frac{\partial}{\partial \mu} + 
\beta(\e, \as) \frac{\partial}{\partial \as} \right) 
K \left( \e, \as \right)~. \nonumber
\eeqa
The anomalous dimension $\gamma_K$ is well known in perturbative QCD. It is 
essentially the anomalous dimension associated with a cusp in a Wilson
line, corresponding here to the hard vertex where the quark and the 
antiquark annihilate \cite{kor}. It is independent of $\e$, and can be 
expanded perturbatively as
\beq
\gamma_K (\as) = \sum_{n = 1}^\infty \gamma_K^{(n)} \left( \frac{\as}{\pi}
\right)^n~,
\label{sergam}
\eeq
with coefficients known to two loops \cite{kotre}
\beq
\gamma_K^{(1)} = 2 C_F \qquad ; \qquad 
\gamma_K^{(2)} = \left( \frac{67}{18} - \zeta(2) \right) C_A C_F -
\frac{5}{9} n_f C_F~.
\label{gamco}
\eeq

Since we are dealing with divergent quantities, it is important to
keep $\e < 0$ throughout. Thus the $\b$ function is defined by 
\beqa
\b(\e,\as) & = & \mu \frac{\partial \as}{\partial \mu} = - 2 \e \as +
\hat{\b} (\as)~, \nonumber \\ \hat{\b} (\as) & = & - \frac{\as^2}{2
\pi} \sum_{n = 0}^\infty b_n \left( \frac{\as}{\pi} \right)^n~.
\label{beta}
\eeqa
At the leading nontrivial order, the solution of \eq{beta} is
\beq
\overline{\alpha}\left(\frac{\mu^2}{\mu_0^2},\as(\mu_0^2),\e\right) =
\as(\mu_0^2) \left[\left(\frac{\mu^2}{\mu_0^2}\right)^\e - \frac{1}{\e}
\left(1 - \left(\frac{\mu^2}{\mu_0^2}\right)^\e \right) \frac{b_0}{4 \pi}
\as(\mu_0^2)\right]^{-1}~,
\label{loalpha}
\eeq 
with $b_0 = (11 C_A - 2 n_f)/3$. Using this form of the running
coupling, and its higher order generalizations, one can solve the
renormalization group equation for the function $G$, as 
\beqa 
G \left(\frac{Q^2}{\mu^2}, \as(\mu^2), \e \right) & = & G \left(-1,
\overline{\alpha} \left(\frac{- Q^2}{\mu^2},\as(\mu^2),\e \right), \e
\right) \nl & + &\frac{1}{2} \int_{- Q^2}^{\mu^2} \frac{d
\lambda^2}{\lambda^2} \gamma_K \left(\overline{\alpha}
\left(\frac{\lambda^2}{\mu^2},\as(\mu^2),\e \right) \right)~,
\label{solG}
\eeqa
where the initial condition was chosen to emphasize that $G$ is real for 
negative $Q^2$.

The second important consequence of dimensional regularization is the
fact that it provides a natural initial condition for \eq{eveq}, in
terms of which one can construct an explicit solution. In fact, one
readily recognizes that each term in the perturbative expansion of
$\Gamma(Q^2)$ must be proportional to a positive integer power of
$(\mu^2/(- Q^2))^\e$. Thus for fixed $\e < 0$ all perturbative
corrections to the form factor vanish in the limit $Q^2 \to 0$, and
one can use
\beq
\Gamma \left(0,\as(\mu^2),\e \right) = 1~.
\label{incon}
\eeq
One finds then
\beqa
& & \hspace{-6mm} \Gamma \left( \frac{Q^2}{\mu^2}, \as(\mu^2), \e \right) ~=~
\exp \left\{ \frac{1}{2} \int_0^{- Q^2} \frac{d \xi^2}{\xi^2} \left[
K \left(\e, \as(\mu^2) \right) \right. \right. \label{sol} \\ 
& + & \hspace{-1mm} \left. \left. G \left(-1, \overline{\alpha} 
\left(\frac{\xi^2}{\mu^2},\as(\mu^2),\e \right), \e \right) 
+ \frac{1}{2} \int_{\xi^2}^{\mu^2} 
\frac{d \lambda^2}{\lambda^2} \gamma_K \left(\overline{\alpha} 
\left(\frac{\lambda^2}{\mu^2},\as(\mu^2), \e \right) \right) \right] 
\right\}~. \nonumber
\eeqa
\eq{sol} can be used directly as a generating function of diagrammatic
results in dimensional regularization; for example, using one--loop
results for the various functions appearing in the exponent, as well
as the one--loop running coupling re--expanded in terms of $\as(Q^2)$,
one reproduces correctly the quartic and cubic poles in $\e$ at ${\cal
O}(\as^2)$. Introducing the two--loop contribution to the anomalous
dimension $\gamma_K$, one matches the double pole as well; the only
singularity requiring a full two--loop computation (the knowledge of
$G^{(2)}$, as well as ${\cal O}(\e^2)$ terms in $G^{(1)}$), is the
single pole, in agreement with the general considerations of
Ref.~\cite{cata}. In Ref.~\cite{us} it was shown that \eq{sol} can be
used to evaluate explicitly the ratio of the timelike to the spacelike
form factor, in terms of a contour integral in the complex plane of
the renormalization scale $\xi^2$. The result (at $\mu^2 = Q^2$) is of
the form
\beqa
\frac{\Gamma \left(1, \as(Q^2), \e \right)}{\Gamma \left(-1, \as(Q^2), \e 
\right)}  \hspace{-4pt} & = & \hspace{-4pt} \exp \left\{ {\rm i} 
\frac{\pi}{2} K \left(\e, \as(Q^2) \right)
+ \frac{{\rm i}}{2} \int_0^\pi d \theta \Bigg[ G \left(-1, \overline{\alpha} 
\left({\rm e}^{{\rm i} \theta}, \as(Q^2), \e \right), \e \right) 
\right. \nl
& - & \left. \hspace{-4pt}
\frac{{\rm i}}{2} \int_0^\theta d \phi ~\gamma_K \left(\overline{\alpha} 
\left({\rm e}^{{\rm i} \phi} ,\as(Q^2), \e \right) \right) \Bigg] \right\}~.
\label{rat}
\eeqa
\eq{rat} proves that the ratio of form factors is given by an infinite 
phase times a finite exponential factor. The derivation of \eq{rat} also 
shows that the ratio is dominated by ultraviolet contributions: in fact, 
it is expressed by a contour integral that never approaches the origin. 
Thus, one may conclude that it is free of power corrections related to the 
Landau pole. In particular, the integrals in \eq{rat} can be evaluated 
explicitly to the desired accuracy in the running coupling; the result is 
of direct phenomenological relevance, since the modulus squared of the ratio
of form factors appears in the resummed partonic Drell--Yan cross 
section~\cite{dysum,MSproc}.

I will now concentrate on \eq{sol}, and examine the structure of the exponent
in more detail. In particular, I will be concerned with the fate of the Landau 
pole, that is expected to make the integration over the infrared region 
ill--defined. It will become apparent that dimensional regularization provides
a solution to this problem. I will however start with an exercise in the 
application of the renormalization group, which displays in a simple context 
the power of dimensional regularization.

\sect{The counterterm function}
\label{count}

Let us begin by considering the ``Coulomb phase'' $K(\e,\as)$. In any minimal 
renormalization scheme, it is a series of poles which can be written 
as\footnote{Notice that $K(\e,\as)$ has only one pole per loop. The leading
pole in the exponent of $\Gamma(Q^2)$, ($\as^n/\e^{n+1}$), is generated at 
each order by the integration of the anomalous dimension $\gamma_K$.}
\beqa
K(\e,\as) & = & \sum_{n = 1}^\infty \frac{K_n(\as)}{\e^n}~, \\
K_n(\as)  & = & \sum_{m = n}^\infty K_n^{(m)} \left( \frac{\as}{\pi} 
\right)^m~. \nonumber
\label{kpole}
\eeqa
In such a scheme, $K(\e,\as)$ has no explicit dependence on the 
renormalization scale $\mu$. The renormalization group equation (\ref{gamk})
then reduces to
\beq
\beta(\e, \as) \frac{\partial}{\partial \as} K \left( \e, \as \right) = 
- \gamma_K (\as)~,
\label{redren}
\eeq
which, given the finiteness of $\gamma_K$ and the form of $\beta (\e, \as)$,
turns into a recursion relation for the perturbative coefficients of $K$.
Explicitly
\beqa
\as \frac{d}{d \as} K_1 (\as) & = & 
\frac{1}{2} \gamma_K (\as)~, \label{recrel} \\
\as \frac{d}{d\as} K_{n + 1} (\as) & = & 
\frac{1}{2} \hat{\beta}(\as) \frac{d}{d \as} K_n (\as)~.
\nonumber
\eeqa
\eq{recrel} expresses the fact that the physics of the function $K(\e,
\as)$ is contained in the residue of the simple pole, $K_1 (\as)$,
which is completely determined by the anomalous dimension $\gamma_K$,
and in turn determines the coefficients of all higher poles. One can
approximate the exact recursion relation by truncating the
four--dimensional $\beta$ function $\hat{\beta}(\as)$ to some chosen
order.

It is useful to collect the terms in $K$ corresponding to leading poles
($(\as/\e)^n$), next--to--leading poles ($\as^{n + 1}/\e^n$), and so on.
This is done by writing
\beqa
K(\e,\as) & = & \sum_{m = 0}^\infty {\cal K}_m (\e, \as)~, \label{serpol} \\
{\cal K}_m (\e, \as) & = & \sum_{n = 1}^\infty K_n^{(n + m)} 
\left(\frac{\as}{\pi}\right)^{n + m} \frac{1}{\e^n}~. \nonumber
\eeqa
According to general arguments, one expects the leading poles in $K$ to be 
completely determined to all orders by the one--loop anomalous dimension and 
the one--loop $\beta$ function; similarly, the $m$--th tower of poles,
${\cal K}_m$ is expected to require the anomalous dimension and the $\beta$ 
function to $m+1$ loops.

In the remainder of this section, the following facts about these towers 
of poles will be established:
\begin{itemize}
\item the recursion relation for the perturbative coefficients $K_n^{(m)}$
can be explicitly solved including {\it all} orders in the $\beta$ function.
\item All the resulting series of poles, ${\cal K}_m(\e, \as)$, can be summed.
\item Upon summation, ${\cal K}_m(\e, \as)$ is an analytic function of
$\as$ and $\e$, {\it regular} as $\e \to 0$ for $m > 0$. The only singularity
at $\e \to 0$ is logarithmic and completely determined by a one loop 
calculation.
\item The finite limits ${\cal K}_m(0, \as)$ ($m > 0$) can be computed for
{\it all} $m$ in terms of the perturbative coefficients of $\beta$ and 
$\gamma_K$. They reconstitute a power series in $\as$.
\end{itemize}
As with most recursion relations, it is useful to start with the simplest 
cases.

\subsection{One loop}
\label{countone}

The first of Eqs.~(\ref{recrel}) is readily solved and yields
\beq
K_1^{(m)} = \frac{1}{2 m} \gamma_K^{(m)}~.
\label{k1}
\eeq
In the second of Eqs.~(\ref{recrel}) one can start by truncating the $\beta$
function at ${\cal O}(\as^2)$, keeping only the one--loop coefficient $b_0$.
Using \eq{k1} one readily finds an expression for the generic coefficient
$K_n^{(m)}$,
\beq
K_n^{(m)} = \frac{1}{2 m} \left(- \frac{b_0}{4} \right)^{n - 1} 
\gamma_K^{(m - n  + 1)}~,
\label{ols}
\eeq
which, as we will verify, is exact for $n = m$. Then
\beq
{\cal K}_0 (\e, \as) = \sum_{n = 1}^\infty K_n^{(n)} \left( 
\frac{\as}{\pi \e} \right)^n = \frac{2 \gamma_K^{(1)}}{b_0} 
\ln \left(1 + \frac{b_0 \as}{4 \pi \e}
\right)~.
\label{klp}
\eeq
Notice that this implies that the resummation of all the leading poles
in the ratio (\ref{rat}) degrades the singularity to a logarithm:
\beq
\left.
\frac{\Gamma \left(1, \as(Q^2), \e \right)}{\Gamma \left(-1, \as(Q^2), \e 
\right)} \right|_{L.P.} = \left(1 + \frac{b_0 \as}{4 \pi \e} \right)^{{\rm i}
\pi \frac{\gamma_K^{(1)}}{b_0}}~.
\label{leadrat}
\eeq
Since all other poles are perturbatively weaker, one might expect that
they should resum to even weaker singularities, or to regular functions. This 
in fact turns out to be the case.

\subsection{Two loops}
\label{counttwo}

To resum next--to--leading poles, one must include the next coefficient of 
the $\beta$ function, $b_1$, as well as $\gamma_K^{(2)}$. \eq{recrel} may 
still be solved, and gives
\beq
K_n^{(m)} = \frac{1}{2 m} \left[ \sum_{k = 0}^{n - 1} \left( 
\begin{array}{c} n - 1 \\ k \end{array} \right) \left(- \frac{b_0}{4} 
\right)^{n - 1 - k} \left(- \frac{b_1}{4} \right)^k 
\gamma_K^{(m - n  - k + 1)} \right]~,
\label{tls}
\eeq
where the sum is further truncated by the requirement that $\gamma_K^{(p)} = 0$
for $p < 1$. One recovers the previous answer, \eq{klp}, for the leading poles.
Next--to--leading poles are given by
\beq
K_{n-1}^{(n)} = \frac{1}{2 n} \left(- \frac{b_0}{4} \right)^{n - 3} 
\left[- \frac{b_0}{4} \gamma_K^{(2)} -
(n - 2) \frac{b_1}{4} \gamma_K^{(1)} \right]~,
\label{knlp}
\eeq
Again, the corresponding series is summable, and yields
\beqa
{\cal K}_1 (\e, \as) & = & \sum_{n = 1}^\infty K_n^{(n + 1)} \left( 
\frac{\as}{\pi} \right)^{n + 1} \frac{1}{\e^n} \nl
& = & \frac{8 \e}{b_0^2} \left(\frac{2 \gamma_K^{(1)} b_1}{b_0} - 
\gamma_K^{(2)} \right) \ln \left(1 + \frac{b_0 \as}{4 \pi \e} \right)
\label{knlps} \\
& - & \frac{\as}{\pi} \left[\frac{2}{b_0} 
\left(\frac{\gamma_K^{(1)} b_1}{b_0} - 
\gamma_K^{(2)} \right) + \frac{2 \gamma_K^{(1)} b_1}{b_0^2}
\frac{4 \pi \e}{b_0 \as + 4 \pi \e} \right]~. \nonumber
\eeqa
The logarithmic singularity receives a new contribution, which, however, 
is supressed by a power of $\e$. Thus ${\cal K}_1$ has a finite limit as 
$\e \to 0$, given by
\beq
{\cal K}_1 (0, \as) = \frac{2 \as}{\pi b_0} \left[\gamma_K^{(2)} -
\frac{\gamma_K^{(1)} b_1}{b_0} \right]~.
\label{fin1}
\eeq

\subsection{All loops}
\label{countall}

Encouraged by the simple form of the answers obtained for ${\cal K}_0$ and
${\cal K}_1$, one sets out to include higher orders in $\hat{\beta} (\as)$.
It turns out that the recursion relation \eq{recrel} is solvable to all 
orders, and the solution is a straightforward generalization of Eqs.
(\ref{ols}) and (\ref{tls}). One finds
\beq
K_n^{(m)} = \frac{1}{2 m} \sum_{p_1, \ldots , p_{n - 1} = 0}^\infty
\left[\prod_{j = 1}^{n - 1} \left(- \frac{b_{p_j}}{4} \right)
\right] \gamma_K^{(m - n + 1 - \sum_{i = 1}^{n - 1} p_i)}~.
\label{als}
\eeq
To identify explicitly the towers of subleading poles, it is useful to
reorganize the sum isolating the perturbative coefficients of
the anomalous dimension $\gamma_K$. This is done by writing \eq{als} as
\beq
K_n^{(m)} = \frac{1}{2 m} \sum_{q_1 = 0}^\infty \gamma_K^{(m - n + 1 - q_1)}
\left\{ \sum_{q_2 = 0}^{q_1} \ldots 
\sum_{q_{n - 1} = 0}^{q_{n - 2}} 
\left[ \prod_{j = 1}^{n - 1}
\left(- \frac{b_{q_j - q_{j + 1}}}{4} \right) 
\right] \right\}~,
\label{ordsol}
\eeq
where the sum over $q_1$ is effectively cut off by the fact that for
$q_1 > m - n + 1$ the summand vanishes.  It is not trivial to
reconstruct the general form of the $m$-th tower of subleading poles
from \eq{ordsol}; however, upon generating a sufficient number of
examples, one recognizes that
\beq
K_{n - m}^{(n)} = \frac{1}{2 n} \left(- \frac{b_0}{4} \right)^{n - m - 1} 
\sum_{p = 1}^{m + 1} C_p (n, m) ~\gamma_K^{(m + 2 - p)}~,
\label{topol}
\eeq
where
\beq
C_p (n, m) = \sum_{k = 0}^{p - 1} \left( \begin{array}{c} n - m - 1 \\ k
\end{array} \right) \left( - \frac{b_0}{4} \right)^{-k} ~{\cal C}_k^{(p)} 
(b_1, \ldots, b_{p - 1})~.
\label{copol}
\eeq
The coefficients ${\cal C}_k^{(p)} (b_1, \ldots, b_{p - 1})$ are
polynomial expressions in the $b_i$'s, given by
\beq
{\cal C}_k^{(p)} (b_1, \ldots, b_{p - 1}) = \sum_{ {\fontsize{8pt}{10pt}
\begin{array}{c} \\[-25pt]
m_i = 1 \\[-8pt] \sum_{i = 1}^k m_i = p - 1
\end{array} } }^{p - 1} \left[
\prod_{i = 1}^k \left( - \frac{b_{m_i}}{4} \right) \right]~.
\label{multic}
\eeq
The key observation concerning Eqs.~(\ref{topol})--(\ref{multic}) is
that the dependence on $n$ is simple, so that summation over $n$ can
explicitly be performed for all $m$. In fact, for fixed $m$,
\eq{topol} can be rewritten as
\beq
K_{n - m}^{(n)} = \frac{1}{2 n} ~\left(- \frac{b_0}{4} \right)^{n - m - 1}~
\sum_{r = 0}^m {\cal D}_r (m) n^r~,
\label{newpol}
\eeq
with easily determined coefficients ${\cal D}_r (m)$. All the series
to be summed are thus simple power series, in fact derivatives of the
logarithmic series which resums the leading poles, up to subtractions
of a finite number of terms. It is easy to verify that the results
obtained for ${\cal K}_m (\e, \as)$ are finite in the limit $\e \to
0$. One can in fact determine the form of this limit explicitly for
all $m$. It is given by polinomial expressions in the coefficients of
the $\beta$ function, similar to Eqs.~(\ref{topol})--(\ref{multic}):
\beq
{\cal K}_m (0, \as) = \frac{1}{m} \left( \frac{\as}{\pi} \right)^m
\frac{2}{b_0} ~\sum_{p = 0}^m B_p (m) ~\gamma_K^{(m + 1 - p)}~,
\label{kfinm}
\eeq
where
\beq
B_p (m) = \sum_{j = 0}^p {\cal B}_j (m, p) \left( 
- \frac{b_0}{2} \right)^{-j}~,
\label{cokfinm}
\eeq
and
\beq
{\cal B}_j (m, p) = \sum_{ {\fontsize{8pt}{10pt}
\begin{array}{c} \\[-25pt]
q_i = 1 \\[-6pt] \sum_{i = 1}^j q_i = p
\end{array} } }^{m} \left[
\prod_{i = 1}^j \left( \frac{b_{q_i}}{2} \right) \right]~.
\label{calb}
\eeq
The most amusing fact here is perhaps that in the limit $\e \to 0$ one
recovers a {\it finite} perturbative expansion in powers of $\as/\pi$.

The conclusion of this exercise is that the counterterm function
$K(\e, \as)$ is completely determined, as an analytic function of
$\e$, $\as$, and the perturbative coefficients of $\beta$ and
$\gamma_K$.  In the limit $\e \to 0$ this function has a logarithmic
singularity, arising from the resummation of the leading poles, and a
contribution which is independent of $\e$ and can be computed as a
perturbative expansion in powers of $\as/\pi$, with coefficients
determined again by the functions $\beta$ and $\gamma_K$. Explicitly
\beq
K (\e, \as) = K_{DIV} (\e, \as) + K_{FIN} (\as) + 
{\cal O}(\e, \e \ln \e)~,
\label{kdivfin}
\eeq
with $K_{DIV} (\e, \as)$ given by \eq{klp}, and
\beqa
K_{FIN} (\as) & = & \sum_{m = 1}^\infty {\cal K}_m (0, \as)  =
\frac{2}{b_0} ~\sum_{m = 1}^\infty \frac{1}{m}
\left( \frac{\as}{\pi} \right)^m {\cal A}_m \label{kfin} \\
{\cal A}_m & = & \sum_{p = 0}^m B_p (m) ~\gamma_K^{(m + 1 - p)}~.
\nonumber
\eeqa
None of the series involved in this computation shows signs of
factorial growth in the coefficients. In the final expression,
\eq{kfin}, the growth of the coefficients is implicit in their
dependence on $b_i$ and $\gamma_K^{(i)}$, the series in \eq{kfin}
being otherwise convergent with a logarithmic behavior.

\sect{One--loop resummation}
\label{onelo}

The success in the computation of the counterterm function $K(\e,
\as)$ is encouraging in view of a more explicit evaluation of the full
form factor.  It is particularly tempting to study the behavior of the
integrals in \eq{sol} in the small--$\xi^2$ region, where one expects
to find obstructions due to the Landau pole, which should however
cancel in the ratio of the timelike to the spacelike factor, \eq{rat},
which is ultraviolet dominated.

A natural place to start is a more precise characterization of the
Landau pole in the context of dimensional regularization. The
one--loop running coupling for finite $\e$ can still be characterized
by a mass scale, defined as the value of the renormalization scale
where the coupling diverges, just as in conventional dimensional
transmutation. In fact, \eq{loalpha} still has a simple pole located at
\beq
\mu^2 = \Lambda^2 \equiv Q^2 \left(1 + \frac{4 \pi \e}{b_0 \as(Q^2)}
\right)^{-1/\e}~,
\label{lapo}
\eeq
and one can use the scale $\Lambda^2$ to define the coupling, as
\beq
\as(Q^2) = \frac{4 \pi \e}{b_0 \left[ \left( \frac{Q^2}{\Lambda^2} 
\right)^\e - 1 \right]}~.
\label{asq}
\eeq
By inspection, Eqs.~(\ref{loalpha}), (\ref{lapo}) and (\ref{asq}) reduce 
to their four--dimensional counterparts as $\e \to 0$. There are however 
two important observations to be made.

\begin{itemize}

\item The running coupling defined by \eq{asq} vanishes as $Q^2 \to 0$
for fixed $\e$, provided ${\bf Re} ~\e < 0$, in agreement with
\eq{loalpha} and with \eq{incon}. This behavior for small $Q^2$ is
simply due to dimensional counting: in fact, since we are keeping the
regulator in place, we are studying the renormalized coupling $g_R =
g_0 \mu^{- \e}$ for fixed bare coupling, in $d > 4$, where it vanishes
as a power of the scale.

\item For general (complex) $\e$, the Landau pole in \eq{lapo} is
located at complex values of the scale $\mu^2$. In particular, for
real negative $\e$, the location of the pole has a nonvanishing
imaginary part provided $\e < - b_0 \as(Q^2)/(4 \pi)$, $\e \neq
-1/n$. Thus, in general, the Landau pole in the dimensionally
regulated exponent of \eq{sol} is {\it not} on the integration
contour. One expects, and I will verify below, that the integrals
appearing in \eq{sol} may be evaluated explicitly in terms of
analytic functions of $\as$ and $\e$. When $\e$ becomes close to $0$,
the pole migrates to the integration contour, and correspondingly the
integral will develop a cut. Nonperturbative information (if any) will
be encoded in the structure of the singularities at the cut, and in
the corresponding higher order generalizations.

\end{itemize}

Armed with these obervations, one can proceed to verify the
integrability of $\log \Gamma(Q^2)$, starting with the simplest
situation, in which only one--loop information is retained. The
integration can be performed using three different methods, which
highlight different features of the exponent.

\subsection{Integration by series}
\label{oneser}

The idea of re--expanding the one--loop running coupling, \eq{loalpha},
in powers of $\as(Q^2)$ is perhaps the first that comes to mind, in
view of its frequent usage in the context of the study of power
corrections.  One writes
\beq
\overline{\alpha}\left(\frac{\mu^2}{\mu_0^2},\as(\mu_0^2),\e\right) =
\left( \frac{\mu_0^2}{\mu^2} \right)^\e \as(\mu_0^2) ~\sum_{n = 0}^\infty 
\left[ \frac{1}{\e} \left( \left( \frac{\mu_0^2}{\mu^2}
\right)^\e - 1 \right) \frac{b_0}{4 \pi} \as (\mu_0^2) \right]^n~,
\label{alser}
\eeq
and one integrates the series term by term, looking for possible
factorial growth in the coefficients of the series obtained after
integration. This would then be interpreted by standard methods as a
sign of nonperturbative contributions. In the present case, changing 
variables in each integral in \eq{sol} as
\beq
\lambda^2 \to z = \left(\frac{\mu^2}{\lambda^2} \right)^\e - 1~,
\label{chvar}
\eeq
one finds that all integrals as easily performed. The integration 
of the anomalous dimension $\gamma_K$ generates terms that are indipendent 
of $\xi$, the integration variable of the outer integral in \eq{sol}. 
These terms are divergent at the lower limit of integration, $\xi^2 = 0$,
but they are cancelled by the $\xi$--independent terms in the expansion
of the counterterm function $K(\e, \as)$, a fact that can be explicitly 
verified to all orders by using the results of \secn{count}. Considering
for simplicity the spacelike form factor (obtained from \eq{sol} by
changing $Q^2 \to - Q^2$), the result
is then
\beqa
\log \Gamma \left(\frac{- Q^2}{\mu^2}, \as(\mu^2), \e \right) & = &
\frac{2}{b_0} ~a(\mu^2) ~\sum_{n = 0}^\infty ~\left\{
\left( \frac{a(\mu^2)}{\e} \right)^n ~\frac{(-1)^{n + 1}}{n + 1}
\right. \label{outser} \\
& \times & \left[ ~-~ \frac{\gamma_K^{(1)}}{2 \e^2} ~\sum_{k = 0}^n 
\, \frac{1}{k + 1} ~\left( \left( 1 - 
\left(\frac{\mu^2}{Q^2} \right)^\e \right)^{k + 1} - ~1 \right) \right. \nl
& - & \left. \left. \frac{G^{(1)} (\e)}{\e} ~\left( \left( 1 - 
\left(\frac{\mu^2}{Q^2} \right)^\e \right)^{n + 1} - ~1 \right) \right]
\right\}~, \nonumber
\eeqa
where I defined
\beq
a(\mu^2) = \frac{b_0}{4 \pi} \as (\mu^2)~,
\label{a}
\eeq
and I wrote the function $G$ perturbatively as
\beq
G( - 1, \overline{\alpha}, \e ) = \sum_{n = 1}^\infty G^{(n)} (\e)
\left( \frac{\overline{\alpha}}{\pi} \right)^n~,
\label{pertg}
\eeq
with $G^{(1)} (\e) = C_F (3 - \e (\zeta(2) - 8))/2 + {\cal O} (\e^2)$.

It is apparent that the series in \eq{outser} is far from
pathological. It does not exhibit any factorial growth, it is
alternating in sign, and definitely has a finite radius of
convergence. In fact, one recognizes a logarithmic series in the terms
with a single sum, whereas the terms in the nested sum are finite
harmonic sums of the kind encountered performing Mellin transforms of
Altarelli--Parisi kernels, which are related to polylogarithms.

Having found that the series arising from the integration of \eq{sol}
is well--behaved, clearly the simplest method to sum it is to perform
the integrals directly, using \eq{loalpha} instead of the Taylor
expansion in \eq{alser}.

\subsection{Analytic expression}
\label{ana}

Direct integration proceeds through steps that are closely related to
the ones outlined in the previous subsection. Through the change of
variable in \eq{chvar} one trivially integrates the one--loop
anomalous dimension, obtaining a logarithm. The logarithm is not
integrable over the scale $\xi^2$ because of the singularity at $\xi^2
= 0$, which is however subtracted by the contribution of the
counterterm function (here one must use for $K(\e, \as)$ the
resummation of the leading poles, \eq{klp}). Using once more the same
change of variables, the second integration can also be
performed. After minor reshuffling with dilogarithm identities, and
inserting the values of $\gamma_K^{(1)}$ and $G^{(1)}$, the result
reads
\beqa
\log \Gamma \left(\frac{- Q^2}{\mu^2}, \as(\mu^2), \e \right) & = &
- \frac{2 C_F}{b_0} ~\left\{ \frac{1}{\e} ~{\rm Li}_2 \left[
\left(\frac{\mu^2}{Q^2} \right)^\e \frac{a(\mu^2)}{a(\mu^2) + \e} \right]
\right. \label{ln1} \\
& & \phantom{aaaaaaa}  \left. - ~C(\e) ~\ln 
\left[1 - \left(\frac{\mu^2}{Q^2} \right)^\e 
\frac{a(\mu^2)}{a(\mu^2) + \e} \right] \right\}~,
\nonumber
\eeqa
where $C(\e) = G^{(1)}(\e)/C_F$.  \eq{ln1} sums the series in \eq{outser},
and has several rather remarkable properties. First, since the form
factor is $RG$ invariant, and I used the exact solution of the
one--loop $RG$ equation for $\overline{\alpha}$ to perform the
integration, \eq{ln1} should be independent of $\mu^2$ to one--loop
accuracy in $\beta(\as)$. This is easily verified, and one obtains
\beqa
& & \hspace{- 1cm} \log \Gamma \left(\frac{- Q^2}{\mu^2}, \as(\mu^2), 
\e \right) ~=~
\log \Gamma \left(- 1, \as(Q^2), \e \right) \label{rgi1} \\
& = & - \frac{2 C_F}{b_0} ~\left\{ \frac{1}{\e} ~{\rm Li}_2 \left[
\frac{a(Q^2)}{a(Q^2) + \e} \right]
+ C(\e) ~\log \left[1 + \frac{a(Q^2)}{\e}\right] \right\}~.
\nonumber
\eeqa
\eq{rgi1}, and its generalization including two--loop effects given in
\secn{twolo}, are the central results of the present paper. The quark
form factor is compactly expressed in terms of a simple analytic
function of the coupling and of $\e$, which is manifestly independent
of the renormalization scale, and can be re--expanded in powers of
$\as(Q^2)$ to generate perturbative results that can be directly
compared with Feynman diagram calculations. In particular, \eq{rgi1}
resums all leading ($\as^n/\e^{2 n}$) and next--to--leading
($\as^n/\e^{2 n - 1}$) infrared and collinear poles in the form
factor.

The second important feature of Eqs.~(\ref{ln1}) and (\ref{rgi1}) is
that it is possible to study the structure of the singularities of the
form factor in the complex $\e$ plane. A detailed analysis of this
question is left for future work. Here I will simply note a few
aspects that are apparent in \eq{rgi1}. First, the Landau singularity,
as expected, appears in \eq{rgi1} as one of the two branching points
of a cut which is shared by both terms. The cut can be taken to run
along the negative real axis in the $\e$ plane (the ``physical''
region), from $\e = - a(Q^2)$ (the Landau singularity, see \eq{lapo})
to $\e = 0$. At the Landau point, both terms in \eq{rgi1} diverge
logarithmically. On the other hand, the limit $\e \to 0$, which is the
physically relevant one, can be studied explicitly; one finds
\beqa
\log \Gamma \left(- 1, \as(Q^2), \e \right) & = &
\frac{2 C_F}{b_0} \Bigg[ - \frac{\zeta(2)}{\e} + \frac{1}{a(Q^2)} +
\left( \frac{1}{a(Q^2)} - \frac{3}{2} \right) \log \left(\frac{a(Q^2)}{\e}
\right) \Bigg. \nl & + & \Bigg. {\cal O} (\e, \e \log\e ) \Bigg]~.
\label{eto01}
\eeqa
\eq{eto01} has interesting features: one observes that resumming all
the leading poles in the logarithm of the form factor softens the
singularity in a manner similar to what happens for the counterterm
function (see \eq{klp}). In this case the leading singularity is a
simple pole, with a residue which does {\it not} depend on the
coupling, an thus not on the scale either. Including higher loops in
$\beta$ corresponds to resumming weaker singularities, thus one might
expect that it would not affect the leading singularity in
\eq{eto01}. In \secn{twolo}, I will verify that this is indeed the
case at two loops. Another interesting feature of \eq{eto01} is the
appearance of a term in the form factor of the form $\exp(c/a(Q^2))$,
which, in the $\e \to 0$ limit, translates into a power behavior of
the form $(Q^2/\Lambda^2)^c$. The term in question in the present case
is not of direct physical significance: it leads to a {\it positive}
fractional power of $Q^2$, and it is tightly connected with the
infrared divergence; in fact, as we will see, it cancels in the ratio
(\ref{rat}), which is the simplest quantity of physical interest that
can be built out of the form factor. It is however interesting that
such a term can arise in this way: one sees that dimensional
regularization and resummation can combine to give well--defined (and
gauge--invariant) power ``corrections'', which might be of physical
interest if the present method can be generalized to construct
infrared safe quantities to the same level of accuracy.

As a final test of the correctness of \eq{rgi1}, one can compute the
ratio $\Gamma(Q^2)/\Gamma(-Q^2)$ in the limit $\e \to 0$.  To the
present accuracy, one finds
\beqa
\log \left[\frac{\Gamma \left(1, \as(Q^2), \e \right)}{\Gamma 
\left(-1, \as(Q^2), \e \right)} \right] & = & 
\frac{2 C_F}{b_0} \left[~{\rm i} \pi \left( 1 + \log \frac{a(Q^2)}{\e}
\right) \right. \label{rat1} \\
& & - \left. \left( \frac{1}{a(Q^2)} - \frac{3}{2} + {\rm i} \pi \right)
\log \left( 1 + {\rm i} \pi a(Q^2) \right)  + {\cal O}(\e) \right]~, 
\nonumber
\eeqa
which agrees with \eq{klp} for the logarithmically divergent term, and
agrees with the results of \cite{us} for the finite contribution,
which was computed there to this same accuracy using \eq{rat}. Notice that
the the term responsible for the power behavior in $Q^2$ has cancelled 
together with the simple pole.

\subsection{Integration over the coupling}
\label{ias}

Generalizing Eqs.~(\ref{ln1}) and (\ref{rgi1}) to include two-- and
higher loop effects would appear to be nontrivial with the techniques
described so far. In fact, even at two loops the $RG$ equation for
$\overline{\alpha}$ cannot be solved in terms of elementary functions
without approximations.  There is, fortunately, no need to do that:
one can change variables according to
\beq
\frac{d \mu}{\mu} = \frac{d \as}{\beta(\as)}~,
\label{chva}
\eeq
{\it i.e.} use the coupling directly as integration variable, without
the need to introduce an explicit expression for $\as (\mu^2)$ in the 
intermediate stages. At one loop
\beq
\frac{d \mu^2}{\mu^2} = - \frac{d \as}{\as} 
\frac{1}{\e + \frac{b_0}{4 \pi} \as}~.
\label{chva1}
\eeq 
One recognizes yet another way to characterize the Landau singularity
in dimensional regularization: it arises because the one--loop $\beta$
function with $\e \neq 0$ does not have a (double) zero at $\as = 0$,
as usual, but has two distinct zeroes, one at the origin and one
located at $\as = - 4 \pi \e/b_0$. This second zero will be on the
integration contour for real negative $\e$, provided $|\e| < b_0
\as(Q^2)/(4 \pi)$, as expected from the earlier calculations. For $\e
<0$, this second zero, where the $\beta$ function is decreasing, is
the one responsible for asymptotic freedom, while at the origin in the
$\as$ plane the $\beta$ function vanishes with positive derivative, as
it would in a QED-like theory, such as $\phi^4$ in $d = 4$.  In fact,
this kind of behavior of dimensionally regularized theories (with the
opposite sign of the $\beta$ function) is familiar in statistical
field theory \cite{paretal}, where dimensional continuation is used,
for example, to study the properties of the (nontrivial) $\phi^4$
theory in $d = 3$ starting from the (trivial) case $d = 4$. In the
present case, one can see from the explicit solution, \eq{loalpha},
that below the critical point $\e = - b_0 \as(Q^2)/(4 \pi)$ the
coupling decreases smoothly to 0, starting from the boundary value
$\as (Q^2)$. Above the critical point, on the other hand (and thus
getting closer to the physical value $\e = 0$), the coupling develops
a Landau pole, so that it becomes impossible to evolve smoothly along
real values of the scale.

Inplementing the change of variables \eq{chva}, one sees that the only
remaining dependence on the scale $\mu^2$ is in the upper limit of
integration of the anomalous dimension integral, which is now $\as
(\mu^2)$, and in the counterterm function $K(\e, \as(\mu^2))$. These
two dipendences must (and do) cancel by $RG$ invariance, in the same
way in which the singularities at $\xi^2 \to 0$ cancelled in the
previous versions of the calculation. At the lower limit of
integration, $\xi^2 = 0$, one must use $\as (0) = 0$, in accordance
with the arguments given above and at the beginning of \secn{onelo};
one then recovers directly \eq{rgi1}.  This third procedure of
integration is clearly, and by far, the most straightforward;
furthermore, it is readily generalizable to to more than one loop, as
I will now show.

\sect{Two--loop resummation}
\label{twolo}

The generalization of the procedure outlined in \secn{ias} to include
higher orders in the $\beta$ function is straightforward. At $l$
loops, $\beta(\as)$ is a polynomial of degree $l + 1$, thus \eq{sol}
will be expressed in terms of a double integral of a combination of
rational function of $\as$, which in general will be computable by
partial fractioning in terms of combinations of polylogarithms. The
result will be expressed in terms of the perturbative coefficients of
the functions $\beta (\as)$, $\gamma_K (\as)$ and $G(1,
\overline{\alpha}, \e)$, which are known to two loops; therefore I
will briefly illustrate the procedure and the results in that case,
where all the ingredients of the final expression are explicitly
known, and the size remains manageable.

At two loops one must use, for each integral over the scale
of the coupling in \eq{sol}, the change of variables
\beq
\frac{d \mu^2}{\mu^2} = - \frac{d \as}{\as} 
\frac{1}{\e + \frac{b_0}{4 \pi} \as + \frac{b_1}{4 \pi^2} \as^2}~.
\label{chva2}
\eeq
Partial fractioning of the denominators is achieved by introducing the
two nontrivial zeroes of the two--loop $\beta$ function, or, for
convenience, the two zeroes of the equation $b_1 \alpha^2 + b_0 \alpha
+ 4 \e = 0$ (with $\alpha = \as/\pi$),
\beq
a_\pm = - \frac{b_0}{2 b_1} \left(1 \pm \sqrt{1 - \frac{16 \e b_1}{b_0^2}}
\right)~.
\label{zeroes}
\eeq
Note that it will be possible to verify that the result reduces to the
correct limit at one loop by using the fact that, as $b_1 \to 0$,
$a_-$ becomes the one--loop zero, $a_- \to - 4 \e/b_0$, while $a_+$
diverges as $a_+ \to - b_0/b_1$.

The cancellation of the dependence on $\as(\mu^2)$ between the counterterm
function $K$ and the anomalous dimension integral takes place as usual.
The final result can be written in the form
\beqa
& & \hspace{-.8cm} \log \Gamma \left(- 1, \as(Q^2), \e \right) =
- \frac{2}{b_1 (a_+ - a_-)} \left\{ \left( G^{(1)}(\e) + 
a_+ G^{(2)}(\e) \right) \log \left(1 - \frac{\as (Q^2)}{\pi a_+} \right)
\right. \nl
& + & 2 ~\left(\gamma_K^{(1)} + a_+ \gamma_K^{(2)} \right) \left[
- \frac{1}{4 \e} {\rm Li}_2 \left( \frac{\as(Q^2)}{\pi a_+} \right) +
\frac{1}{2 b_1 a_+ (a_+ - a_-)} \log^2 \left(1 - \frac{\as(Q^2)}{\pi a_+}
\right) \right. \nl
& & - ~\frac{1}{b_1 a_- (a_+ - a_-)} \left( {\rm Li}_2 \left( 
\frac{\pi a_+ - \as(Q^2)}{\pi (a_+ - a_-)} \right) - 
{\rm Li}_2 \left( \frac{a_+}{a_+ - a_-} \right) \right. \nl
& & \left. \left. \left. \hspace {-1mm} + ~\log 
\left(1 - \frac{\as(Q^2)}{\pi a_+} \right) 
\log \left(\frac{\as(Q^2) - \pi a_-}{\pi  (a_+ - a_-)} \right) \right)
\right] \right\} ~+ ~\left( a_+ \leftrightarrow a_- \right)~.
\label{rgi2}
\eeqa
It is fairly straightforward to check that \eq{rgi2} reduces to
\eq{rgi1} when one takes $\{b_1, \gamma_K^{(2)}, G^{(2)}(\e) \} \to
0$.  On the other hand, the analytic structure of \eq{rgi2} is
considerably more complicated than the one briefly discussed at one
loop. In particular, the presence of a second nontrivial zero in
$\beta(\as)$ intoduces a new branch point, so that the structure of
cuts becomes more complicated.  One does not expect, however, the
behavior physical quantities related to the form factor to be
qualitatively affected by the inclusion of two--loop effects, since
these are understood to modify the running of the coupling by an
amount which is a logarithmically suppressed and smooth funcion of the
energy.

It is interesting to study the behavior of the two--loop resummed form
factor, \eq{rgi2}, in the neighborhood of the ``physical'' limit, $\e
\to 0$. One finds that, as expected, the simple pole in the logarithm
of the form factor given in \eq{eto01} is not affected by the
inclusion of two--loop effects. The logarithmic singularities,
however, are enhanced to $\log^2$ strength, and also the constant
terms in the limit $\e \to 0$ receive new nontrivial
contributions. One finds
\beqa
& & \hspace{-8mm} \log \Gamma \left(- 1, \as(Q^2), \e \right) ~=~
- \frac{\gamma_K^{(1)}}{b_0} \frac{\zeta(2)}{\e} +
\frac{2 b_1 \gamma_K^{(1)}}{b_0^3} \log^2 \left[ \frac{4 \pi \e}{b_0 \as(Q^2)}
\left(1 + \frac{\as(Q^2) b_1}{\pi b_0} \right) \right] \nl
& & + ~\frac{2}{b_0} \left( G_0^{(1)} + \frac{2 \gamma_K^{(2)}}{b_0} -
\frac{2 \pi \gamma_K^{(1)}}{\as(Q^2) b_0} - 
\frac{2 b_1 \gamma_K^{(1)}}{b_0^2} \right) 
\log \left[ \frac{4 \pi \e}{b_0 \as(Q^2)}
\left(1 + \frac{\as(Q^2) b_1}{\pi b_0} \right) \right] \nl
& & + ~\frac{4 \gamma_K^{(2)}}{b_0^2} {\rm Li}_2 \left(
\frac{\as(Q^2) b_1}{\pi b_0 + \as(Q^2) b_1} \right)
+ ~\left( \frac{4 \pi \gamma_K^{(2)}}{\as(Q^2) b_0 b_1} -
\frac{2 G_0^{(2)}}{b_1} \right) \log \left(1 + \frac{\as(Q^2) b_1}{\pi b_0}
\right) \nl
& & + ~\frac{4 \pi \gamma_K^{(1)}}{\as(Q^2) b_0^2}
+ \left(1 - 2 \zeta(2) \right) \frac{4 b_1 \gamma_K^{(1)}}{b_0^3} -
\left(1 - \zeta(2) \right) \frac{4 \gamma_K^{(2)}}{b_0^2} +
{\cal O} \left( \e, \e \log \e \right)~,
\label{eto02}
\eeqa
where $G^{(i)}_0 = \lim_{\e \to 0} G^{(i)} (\e)$. Also in this limit, 
it is easy to check that taking $\{b_1, \gamma_K^{(2)}, G^{(2)}(\e) \} 
\to 0$ one gets back \eq{eto01}.

Although it is not proved here, one can conjecture rather confidently
that the simple pole in $\log \Gamma(Q^2)$ is completely determined at
one loop, and receives no corrections from higher orders in
$\beta$. The presence of the pole is in fact tightly connected with
the one--loop zero of the $\beta$ function, which is located at a
distance ${\cal O}(\e)$ from the origin. All higher--order zeroes have
locations in the $\as$ plane that are independent of $\e$ in the small
$\e$ limit, thus they are not expected to introduce further singular
behavior in this region. Logarithmic singularities, on the other hand,
arise both from the anomalous dimension integral and from the
integration of the function $G$, thus one must expect that they get
higher order corrections.

\sect{Summary and perspectives}
\label{sumpe}

I have discussed the quark form factor in the context of dimensionally
regulated perturbative QCD. Building upon the results of
Ref.~\cite{us}, as well as upon the understanding developed previously
by many authors \cite{suda}--\cite{sen}, I have been able to derive
resummed analytic expression for the form factor that are independent
of the choice of renormalization scale, and can be systematically
improved upon by including higher orders in the QCD $\beta$ function.

The main ingredient in the derivation is the consistent usage of
dimensional regularization as analytic continuation to general complex
values of the space--time dimension, $d = 4 - 2 \e$. It turns out that
this continuation does not only regulate ultraviolet (as well as
infrared) singularities in Feynman integrals: it also regulates
singularities such as the Landau pole, which arise upon resummation of
classes of Feynman diagrams, and are usually interpreted as signals of
nonperturbative physics. How this happens is well understood, and this
understanding has been applied as a technical tool in statistical
field theory for a long time: for $d > 4$, the QCD $\beta$ function
develops an asymptotically free fixed point at a distance ${\cal O}
(\e)$ from the origin in the $\as$ plane, while the origin itself
becomes {\it infrared} free; the coupling vanishes like a power of the
renormalization scale, and it does so smoothly for sufficiently large
$d$; when $d$ gets close to the physical value $d = 4$ the coupling
develops a Landau pole for real values of the scale; this pole is then
found on the integration contour of resummed expressions for QCD
amplitudes, which, as a consequence, develop cuts with computable
branching points and discontinuities.

The versatility of dimensional regularization was exploited here also
to derive compact resummed analytic expressions for the counterterm
function $K(\e , \as)$: the fact that the renormalization group
equation for a pure counterterm (with a finite anomalous dimension)
turns into a recursion relation is well known; the fact that for the
quark form factor the recursion relation is completely solvable to all
orders in $\beta$, and the resulting towers of poles can be summed,
provided a strong motivation to pursue the calculation for the
complete form factor.

The calculations presented in this paper are not directly applicable
to physical processes: the paper, after all, deals with a divergent
quantity\footnote{The modulus squared of \eq{rat1} is an exception, as
it enters directly the resummed Drell--Yan cross section; it was,
however, computed already in Ref.~\cite{us}}. There are however
several lines of enquiry that are opened by these results, and I
believe are worth pursuing.

First of all, it would be of great interest to extend results such as
\eq{rgi1} and \eq{rgi2} to more general QCD amplitudes, in particular
to amplitudes with more than two colored particles. This is not
straightforward, because more complicated amplitudes have a nontrivial
tensor structure in color space, and evolution equations like
\eq{eveq} correspondingly turn into matrix equations. It should be
emphasized however that much of the necessary technology has already
been developed in the context of the resummation of large--$x$
logarithms for the production of heavy quark pairs and other colored
final states in hadronic collisions \cite{ttbar}.  In that case one
resums large corrections due to the presence of $+$ distributions at
each order in perturbation theory, arising from the imperfect
cancellation of real and virtual contributions. In the present case
one would be interested in purely virtual corrections contributing to
the cross section with terms proportional to $\delta$ functions. I
believe that a generalization of the present results in this direction
should be possible, and work to achieve it is under way.

A more ambitious goal would be to attempt the construction of {\it
finite}, physical quantities to the same accuracy achieved here for
the form factor.  This would require controlling the
infrared/collinear singularities of dimensionally regulated real
emission amplitudes to all orders. While this is not likely to be
possible in a general multi--scale process, progress might conceivably
be made in the simplest cases, such as the inclusive Drell-Yan cross
section, by introducing suitable approximations to the
$d$--dimensional cross section.

On a more speculative note, one may wonder whether the analytic
structure uncovered in the present paper may shed some light on
nonperturbative features of QCD amplitudes. It was already noticed in
the comments to \eq{eto01} that in the neighborhood of $\e = 0$ one
finds terms that translate into powers of the ratio
$Q^2/\Lambda^2$. More generally, the analytic features of the form
factor near the Landau cut might provide suggestions to build models
for the nonperturbative behavior of simple QCD amplitudes, as well as
for the behavior of the coupling itself, much as renormalon
calculations did \cite{bene,reno}.

In the context of finite order perturbative QCD calculations, the
existence of genuinely $RG$--invariant expressions such as
Eqs.~(\ref{rgi1}) and (\ref{rgi2}) might prove useful as a model for
the renormalization scale dependence of more complicated QCD
amplitudes. Not surprisingly, the results for the form factor are
naturally expressed in terms of $\as(Q^2)$, since $Q^2$ is the only
perturbative scale in the amplitude. One can however go back to
equations such as (\ref{ln1}), choose a different renormalization
scale, say $\mu = Q/2$, re--expand the form factor to a finite
perturbative order, and study the variations of the answer with $\mu$.

Finally, it should be mentioned that large corrections to perturbative
QCD amplitudes due to analytic continuation (the ``$\pi^2$ terms'') do
not arise only in quark--initiated process. A case in point is Higgs
production via gluon fusion, in an effective theory in which the top
quark has been integrated out~\cite{dawspi}. Also in that case the
present techniques may well prove useful and provide a very accurate
and rigorous evaluation of the impact of these terms to all orders in
the coupling.

In summary, I believe that the results presented here should be of
interest for the study of several aspects, both formal and practical,
of perturbative QCD. Much work remains to be done.

\vskip 1cm

{\large {\bf {Acknowledgements}}}

It is a pleasure to thank George Sterman for getting me started into this 
field a long time ago, as well as for early discussions of this work.
I would also like to thank Michele Caselle for his insights concerning
the use of dimensional regularization in statistical field theory.

\end{document}